\documentclass[aps,prl,twocolumn,superscriptaddress, 10pt]{revtex4-1} 
\usepackage{graphicx}
\usepackage{amssymb, amsmath, amsthm, amsfonts, bm, mathrsfs, wasysym, txfonts, bbm, enumerate}
\usepackage[usenames,dvipsnames]{xcolor}
\usepackage{slashed}

\begin{document}

\title{{\large Dark fermions from the Standard Model via spin-charge separation}}

\author{Chi Xiong}
\email[]{xiongchi@ntu.edu.sg}
\affiliation{Institute of Advanced Studies, Nanyang Technological University, Singapore 639673, Singapore }
\begin{abstract}
We study a new composite scenario of the lepton sector in the Standard Model by a de-gauging procedure called spin-charge separation and propose that leptons are bound states of some neutral fermions and Higgs bosons. Continuing this procedure we may obtain more fundamental dark fermions. They become the physical leptons by acquiring both charges and masses from some Higgs fields.
\end{abstract}

\maketitle              

In condensed matter physics, spin-charge separation is associated with the resonating valence bond (RVB) theory \cite{Anderson73_87} of high-temperature superconductivity. Electrons in some materials can be considered as ``composite" ones and in the deconfinement phase, split into independent excitations called spinons and chargons (or holons) which carry the spin and the electric charge, respectively. Many elaborations of this idea followed in the studies of high $T_c$ superconductors (see e.g. \cite{Wen:2006} for a review), and so did the experimental observations and computer simulations --- the first direct observations of spinons and holons was reported in \cite{Shen:2006}; Simulations on spin-charge separation via quantum computing has been performed in  \cite{Kwek:2011}. As an example we look at the decomposition in the slave-boson formalism of the t-J model  \cite{Barnes:1976}, which describes the low-energy physics of the high-temperature cuprates,
\begin{equation} \label{cfb}
 c^{\dagger}_{i \sigma} = f^{\dagger}_{i \sigma} b_i 
\end{equation}
where $c^{\dagger}_{i \sigma}$ is the electron operator, and the operators $f^{\dagger}_{i \sigma}$ creates a chargeless spin-$\frac{1}{2}$ fermion state (spinon) and $b_i $ creates a charged spin-0 boson state (holon). The d-wave high-$T_c$ superconducting phase appears when holons condense, $\left\langle b_i^{\dagger} b_i \right\rangle \neq 0$. This spin-charge separation occurs in particular material environments, not for isolated electrons. Nevertheless, in particle physics composite models of (fundamental) particles may allow such a phenomenon. We first look at a couple of unification models from the spin-charge separation point of view.   

In the Pati-Salam model \cite{Pati-Salam-74} based on the gauge group $SU(2)_L \times SU(2)_R \times SU(4)_c$, and the grand unification theory based on the gauge group $SO(10)$ \cite{Fritzsch:1974}, right-handed neutrinos are introduced, and the usual color group $SU(3)_c$ is extended to the group $SU(4)_c$ and the lepton number is interpreted as the fourth color
\begin{equation} \label{FB}
\Psi_{L, R} = \left( \begin{array}{cccc}
u_1 & u_2 & u_3 & \nu_e \\ 
d_1 & d_2 & d_3 & e^{-} \\ 
c_1 & c_2 & c_3 & \nu_{\mu} \\ 
s_1 & s_2 & s_3 & \mu^{-} 
\end{array} \right)_{L, R} = \left( \begin{array}{c}
\mathcal{F}_1 \\ 
\mathcal{F}_2 \\ 
\mathcal{F}_3 \\ 
\mathcal{F}_4
\end{array}  \right)_{L, R} ~\otimes ~ (\mathcal{B}_1, \mathcal{B}_2,  \mathcal{B}_3,  \mathcal{B}_4)
\end{equation}
where $\mathcal{F}$ are fermions carrying the spin of $\Psi$ while $\mathcal{B}$ are scalars carrying the (color) charge of $\Psi$. As pointed out in \cite{Pati-Salam-74}, it is attractive to consider $\Psi$ as {\it composite} particles and $(\mathcal{F}, \mathcal{B})$ as  {\it fundamental} ones. Similar ideas has been used in some preon models, e.g., the ``haplon" model \cite{Fritzsch:1981} which consider quarks and leptons as bound states of some more fundamental particles (preons) called haplons. The haplons are a weak-$SU(2)$ doublet of colorless fermions $(\alpha, \beta)$, and a quartet of scalars $(x^1, x^2, x^3, y)$ with $y$ carrying the fourth color. The first generation fermions read
\begin{equation} \label{haplon}
\nu = (\alpha y), ~e^{-} = (\beta y), ~ u = (\alpha x), ~d = (\beta x). 
\end{equation}
which has the same spin-charge separation pattern $\Psi = \mathcal{F} \mathcal{B}$  as in Eqs. (\ref{cfb}) and ( \ref{FB}). 

If spin-charge separation can be realized in both electric and color charge cases, it should be realized for the weak charge case as well and we will show how this could be proceeded. However, we will not only be doing the generalization, but also pushing the idea to extremes --- what about if we separate all the charges of quarks and leptons from their spins? We might obtain some fundamental fermions which probably interact with each other only gravitationally.        The spin-charge separation might happen step by step and we call it a ``de-gauging" process of quarks and leptons, and the resultant spinon ``dark fermions", since at some stage they might be proper candidates for dark matter. 

First we use the sigma model as a simple example to demonstrate the idea. The Lagrangian reads
\begin{eqnarray}
\mathcal{L}_{\sigma} &=& \bar{\psi}_L i \slashed{\partial} \psi_L + \bar{\psi}_R i \slashed{\partial} \psi_R + \frac{1}{4} \textrm{Tr} \,(\partial \Sigma \cdot \partial \Sigma^{\dagger}) \cr
&-& V(\Sigma, \Sigma^{\dagger}) - y ( \bar{\psi}_L \Sigma \psi_R +  \bar{\psi}_R \Sigma^{\dagger} \psi_L ), 
\end{eqnarray}
which is invariant under the $SU(2)_L \times SU(2)_R$ transformations $\psi_{L,R} \rightarrow \mathcal{A}_{L, R} \, \psi_{L,R}, ~\Sigma \rightarrow \mathcal{A}_L \Sigma \mathcal{A}_R^{\dagger}, ~ \mathcal{A}_{L, R} \in SU(2)_{L,R}$. 
The matrix field $\Sigma$ can be parametrized by either  $\sigma$ and the ``pions" $\vec{\pi}$ or the polar variables $\eta$ and $\vec{\zeta}$ (exponential parametrization)
\begin{equation} \label{scs_Sigma}
\Sigma = \sigma (x) + i \vec{\tau} \cdot \vec{\pi} (x) =  [v + \eta(x)] U(x), ~~~U= e^{i \vec{\tau} \cdot \vec{\zeta}(x)/v },
\end{equation}
where $v$ is the vacuum expectation value of $\sigma$. In terms of $\eta$ and $U$, the $SU(2)_L \times SU(2)_R$ transformation laws become  
$\eta \rightarrow \eta, ~~ U \rightarrow \mathcal{A}_L U \mathcal{A}_R^{\dagger}$. Therefore the $U$ field inherits the transformation property of $\Sigma$ while the $\eta$ field is ``neutralized". 
From spin-charge separation point of view,  we consider the $\eta$ field as a ``spinon", which in this case is a real scalar with $0$-spin and the $U$ field as the chargon, respectively, plus the second parametrization in (\ref{scs_Sigma}) as a spin-charge separation for the $\Sigma$ field.   

We then consider vector-like theories which the spinons will be Majorana type. Starting with an Abelian gauge theory for a Dirac fermion $\Psi_D$ and a gauge field $A_\mu$. 
\begin{equation}
\mathcal{L}_{V} = \bar{\Psi}_D ( i \slashed{\partial} + e \slashed{A} ) \Psi_D - (1/4) ~F^2, 
\end{equation}
Like the complex scalar field $\Sigma$, the Dirac fermions can be considered as a combination of two Majorana fermions $\Psi_D  = \Psi_M^1 + i \Psi_M^2$. The spin-charge separation suggests $\Psi = \Psi_M \mathcal{B} $, where the ``spinon" $\Psi_M$ is a Majorana spinor. The coupling $A_\mu J^\mu$ then vanishes as $ J^\mu \propto \overline{\Psi}_M \gamma^\mu \Psi_M= 0$,  hence the Majorana spinon $\Psi_M$ decouples from the gauge field $A_\mu$. The interaction of $\Psi_M$ to the chargon $\mathcal{B}$ vanishes as well since it is proportional to $ (\mathcal{B} \partial_\mu \mathcal{B}^* - \mathcal{B}^* \partial_\mu \mathcal{B}) ~ \overline{\Psi}_M \gamma^\mu \Psi_M$.
Thus we obtain a complete spin-charge separation (de-gauging) of a Dirac fermion for this vectorial theory if the spinon is taken to be the Majorana-type. 

For chiral theories like the electroweak theory in the Standard Model, it is more complicated since in the (3+1)-dimensional Minkowski spacetime spinors cannot satisfy both Weyl and Majorana conditions. It turns out that the spin-charge separation has to be performed in an asymmetric way (with respect to the left-right symmetry). On the other hand, the existence of neutrinos seems to suggest that they are natural candidates for the spinons from the electromagnetic-charge-spin separation \cite{XC_15}. As the neutrinos and electrons form an $SU(2)_L$ doublet in the Standard Model, we consider their weak-charge-spin separation together. If the neutrinos are closely related to the spinons, what will be chargons? Noticing that left-handed lepton doublet $l_L \equiv ( \nu, e^-)^T_L$ and the tilde Higgs doublet $\tilde{\Phi}$ both transform as $(1, 2, -1/2)$ under the symmetry group $SU(3)_C \times SU(2)_L \times U(1)_Y$, one may conjecture that, for the first generation,
\begin{equation} \label{guess}
 \left( \begin{array}{c}
\nu \\ 
e^-
\end{array} \right)_L \sim  \tilde{\Phi} \otimes \mathcal{F}_L = \left( \begin{array}{c}
~\phi^{0*} \mathcal{F}_L \\ 
- \phi^- \mathcal{F}_L 
\end{array}  \right)
\end{equation}
where the neutral spinor $\mathcal{F}_L$ is a dark fermion that we look for, and the Higgs fields are the chargons. A precise expression of (\ref{guess}) will be calculated and given later.  For the right-handed lepton singlets of $SU(2)_L$, we need extra scalar fields $\chi^-$  and $\chi^0$ for their chargon parts
\begin{equation}
e^{-}_R \sim \chi^- \mathcal{F}_R, ~~~\nu_R  \sim \chi^0 \mathcal{F}_R. 
\end{equation}
where the spinor $\mathcal{F}_R$ is another dark fermion. From the composite model point of view, this is different from the Haplon model in comparison with (\ref{haplon}).  We will discuss the composite features such as binding forces later.   

We know that for compact gauge groups it is always possible to choose the unitary gauge and write the Higgs doublet in a polar form or exponential parametrization similar to (\ref{scs_Sigma})
\begin{equation} \label{Ugauge}
\Phi (x)  = U^{-1} (\xi) \left( \begin{array}{c}
0 \\ 
\frac{h(x)}{\sqrt{2}}
\end{array}  \right), ~~~~~~U(\xi) = e^{-i \vec{\tau} \cdot \vec{\xi}/v}.
\end{equation} 
As mentioned in the sigma-model example, this is the spin-charge separation form for the Higgs doublet and the field $h(x)$ plays the role of spin-0 spinon.  With $U(\xi)$ a field redefinition for the lepton doublet can be made,
\begin{equation} \label{redef}
l_L = \left( \begin{array}{c}
\nu \\ 
e^{-}
\end{array}  \right)_L = U^{-1} \left( \begin{array}{c}
\nu^{'} \\ 
e^{'-}
\end{array}  \right)_L, ~~~~~~e^{-}_R = e^{' - }_R, ~~\nu_{R} = \nu^{'}_{R}. 
\end{equation}
It is not difficult to find out
\begin{equation}
\left( \begin{array}{c}
\nu \\ 
e^{-}
\end{array}  \right)_L = \frac{1}{h} \left( \begin{array}{c}
\phi^{0*}\nu^{'} + \phi^{+} e^{'-} \\ 
- \phi^{-} \nu^{'} + \phi^{0} e^{'-}
\end{array}  \right)_L  
\end{equation}
or in a more compact form
\begin{equation}
l_L = \frac{1}{h} \, (\tilde{\Phi} \otimes  \nu^{'}_L + \Phi \otimes e^{'-}_L)
\end{equation}
This is an interesting relation whose first term on the right-hand side resembles our guess (\ref{guess}). Noticing that fields $ \nu^{'}_L$ and $e^{'-}_L$ now has the same quantum number as the $ \nu^{'}_R$ and $e^{'-}_R$, respectively, one may realize that the separation of the weak charge makes the right-handed and left-handed fermions symmetric. This is of course not a surprise and reflects the consistency of our spin-charge separation procedure.

Let us do spin-charge separation for the electron fields  $e^{'-}_L$  and  $e^{'-}_R$ in a left-right symmetric way
\begin{equation}
e^{'-}_L = \chi^{-}  \,\mathcal{F}_L~, ~~~~~~~ e^{'-}_R =  \chi^{-} \, \mathcal{F}_R~,
\end{equation}
where the fermion fields $\mathcal{F}_L, \mathcal{F}_R$ are the dark fermions we expected, and the extra scalar $\chi^{-}$ has been made dimensionless by rescaling.  For the neutrino fields $ \nu^{'}_L$ and $ \nu^{'}_R$, we assume that they have the same spinon part as the electron fields,
\begin{equation}
\nu^{'}_L =  \chi^{0} \,\mathcal{F}_L~, ~~~~~~~ \nu^{'}_R =  \chi^{0} \, \mathcal{F}_R~,
\end{equation}
Therefore the spin-charge separation for the original fields of the electron and electron neutrino is
\begin{eqnarray}
\nu_L &=& \frac{1}{h} \left( \phi^{0*} + \phi^{+} \chi^{-}  \right) \, \mathcal{F}_L~, ~~~~~~\nu_R =  \chi^{0} \, \mathcal{F}_R \cr
e^{-}_L &=& \frac{1}{h} \left( - \phi^{-} + \phi^{0} \chi^{-}  \right) \, \mathcal{F}_L~, ~~~~~ e^{-}_R = \chi^{-} \, \mathcal{F}_R
\end{eqnarray} 
Note that we have introduced extra scalars $\chi^-, \chi^0$ for the electric-charge-spin separation. These scalars are important in distinguishing electrons and neutrinos. However, it depends on a detailed study of the binding forces and dynamics of the bound states.  We close by discussing a few more issues:

1). Other generations: So far we have shown that the first generation leptons can be considered as bound states of some Higgs fields and the dark fermion. A naive generalization would be that there are other generations of dark fermions. This solution, however, does not fully take the advantage of the bound states. For example, the radical excitations of a bound state might provide other generations \cite{Visnjic}. Another possibility is that the second and the third generation might be the bound states of the first generation and one and two Higgs fields \cite{Derman:1980rr} or other fields \cite{Fritzsch:1981}, e.g. $\mu \sim [L_1h], ~\tau \sim [L_1 h h]$. There also could be fermion-string bound-states in which {\it chiral} fermion zero-modes are trapped in a vortex configuration of the scalar fields \cite{XC_15, XC};

2). Binding forces: In the high-temperature cuprate superconductivity case, the binding force is provided by some emergent $U(1)$-gauge field \cite{Wen:2006}. In the preon models, some hyper-color fields are introduced (e.g. the QHD gluons in the haplon model \cite{Fritzsch:1981}). 
We do not follow these options, instead, assume that some Yukawa type-interaction provide such binding force. This is because we do not want the dark fermions to interact with some new gauge fields, otherwise our spin-charge separation scenario would be ruined. Note that gauge symmetry is not the only way to introduce interactions between matter fields. Supersymmetry can bring fermion-boson interactions \cite{WB}. The Wess-Zumino model provides an interacting theory which only involves scalars and spinors. Interestingly, it can be considered as Majorana fermions interacting with complex scalars. As a matter of fact, it is natural for the supersymmetry to play a role in our ``de-gauging" procedure in which what are finally left are spacetime symmetries. 

3). Other theories: The ``de-gauging" procedure of the leptons in the Standard Model can be generalized to other gauge theories with Higgs mechanism. As we demonstrated above for the Standard Model, the polar form or exponential parametrization of the Higgs (\ref{Ugauge}) can always be reached since there always exists a unitary gauge in the compact gauge group cases. One then can make field-decomposition as in (\ref{redef}) and identify the chargon and the spinons. This procedure depends on the symmetry-breaking pattern and as it goes on, more fundamental fermions (in the sense that they carry less charges) will emerge and symbolically written as $\mathcal{F} \sim \Psi / H$, where $H$ stands for the Higgs sector of the theory.

4). Composite bosons: A complete fermionic theory is possible if the chargons are also composite ones. For example in \cite{Hill_15} the Higgs boson is composed of neutrinos. Gauge bosons could be composite ones as well, e.g. the composite W-bosons in the haplon model \cite{Fritzsch:1981}. 

5). Quarks and color charges:  We will not address in this Letter the spin-charge separation of quarks and non-Abelian gauge fields.  These are more complicated issues (however see \cite{Faddeev:2006, XC} for the non-Abelian gauge field cases) and we leave them for future investigations.   

It is interesting to look at the masses and charges of the Standard Model leptons from the dark fermion point of view --- it seems that the Higgs boson not only gives mass to leptons, but also give them charges.

\vspace{1cm}

{\it Acknowledgment}: I thank Peter Minkowski, Harald Fritzsch and HweeBoon Low for valuable discussions. This work has been supported by the Institute of Advanced Studies, Nanyang Technological University, Singapore.


\end{document}